\documentclass[final,epsfig,amsfont]{article}
\usepackage{latexsym}
\usepackage[dvips]{graphicx}
\newcommand{\veps}{\varepsilon}
\begin{document}

\title{On the statistical distribution of first--return times of balls and cylinders
in chaotic systems}

\date{}
\author{G.Mantica\thanks{International Center for
Non-linear and Complex Systems, Universit\`a dell'Insubria, Via
Vallegio 11, Como, and CNISM, unit\`a di Como, I.N.F.N. sezione di
Milano, Italy. E-mail: $<$giorgio@uninsubria.it$>$.}, and
S.Vaienti
\thanks{Centre de Physique Th\'eorique, UMR 6207, CNRS, Luminy
Case 907, F-13288 Marseille Cedex 9, and Universities of
Aix-Marseille I, II and Toulon-Var. F\'ed\'eration de Recherche
des Unit\'es de Math\'ematiques de Marseille, France. E-mail:
$<$vaienti@cpt.univ-mrs.fr$>$.}}

\maketitle

\begin{abstract}
We study returns in dynamical systems: when a set of points,
initially populating a prescribed region, swarms around phase
space according to a deterministic rule of motion, we say that the
return of the set occurs at the earliest moment when one of these
points comes back to the original region. We describe the
statistical distribution of these ``first--return times'' in
various settings: when phase space is composed of sequences of
symbols from a finite alphabet (with application for instance to
biological problems) and when phase space is a one and a
two-dimensional manifold. Specifically, we consider Bernoulli
shifts, expanding maps of the interval and linear automorphisms of
the two dimensional torus. We derive relations linking these
statistics with R\'enyi entropies and Lyapunov exponents.
\end{abstract}
\section{Introduction}

In this paper we investigate a phenomenon of vast relevance in
physics: returns. Whenever a system evolves according to a
deterministic, or even a probabilistic law, along the course of
time it may pass close to points previously visited. We then speak
of returns and of return times. Of course, this concept can be
made precise and rigorous: this has been done since the beginning
of the theory of dynamical system, where Poincar\'e
theorem---establishing that in measure preserving systems returns
happen almost surely---is probably the first and certainly the
most celebrated result. Passing via Kac theorem and coming to
recent years, mathematical investigation has flourished and
produced beautiful results relating the statistics of return
times, {\em i.e.} the collective counting of these values, to more
conventional dynamical indicators, such as generalized dimensions
of invariant measures, Lyapunov exponents and the like. This paper
continues in this ongoing investigation, with a special character:
rather than presenting a single, thoroughly investigated result,
we attempt to provide a heuristic, global picture of the dynamical
phenomena that are at work. This picture will be confirmed by
numerical experiments. Rigor and detailed proofs will be the
matter for successive publications.

There exist many different alternatives when defining returns: in
this paper, we make specific reference to what is called { the
first return} of a (measurable) subset $A$ of a compact metric
space, endowed with a measure $\mu$ defined on the Borel sigma algebra
${\cal A}$. Motion on $X$ is effected by the action of a
transformation $T$ that preserves the measure $\mu$. We thereby
define the {\em time of first return  of $A$} into itself as
\begin{equation} \label{deftc}
\tau({A})\; :=\; \min\{k > 0 \mbox{ s.t. } T^k A \cap A  \,\neq
\,\emptyset\}\;.
\end{equation}
As anticipated in the abstract, this is the time of the earliest
return to $A$ of one of its points:
\begin{equation} \label{deftc2}
\tau({A})\; =\; \min\{k > 0 \mbox{ s.t. } \exists x \in A; \,T^k
(x) \in A \}.
\end{equation}

 Two choices will be made for $A$. Firstly, $A$ will be a ball
of radius $\varepsilon$, centered at a point $x \in X$:
$B_\varepsilon(x)$. Secondly, $A$ will be a dynamically generated
cylinder. Let us suppose that $\mathcal{C}$ is a finite partition
of $X$ and take the n-join, $\mathcal{C}^n :=
\vee_{i=0}^{n-1}T^{-i}\mathcal{C}$. We call cylinder of length $ n
$ around $ x \in X $, denoted with $C_n(x)$, the unique element of
$ \mathcal{C}^n $ containing $x$. The statistics of return times
are then defined by the collective counting, over different sets,
of the values $\tau(A)$. The distribution of return times,
$p(\varepsilon,k)$ is
\begin{equation}
 \label{defstat}
 p(\varepsilon,k) := \mu(\{ x \in X \; \mbox{ s.t } \;
  \tau({B_\varepsilon(x)})  = k \});
\end{equation}
similarly, $p(n,k)$ is defined replacing $B_\varepsilon(x)$ by
$C_n(x)$. They measure the fraction of points in the space $X$
whose neighborhood (whether a ball of radius $\veps$, or a
cylinder of length $n$) first returns to itself after $k$
iterations of the map. We shall also consider the cumulative
distributions (integrated statistics) $P(\varepsilon,k)$,
\begin{equation}
 \label{defstat2}
 P(\varepsilon,k) :=  \sum_{j=1}^k p(\varepsilon,j)
\end{equation}
and $P(n,k)$. Our aim will be to study the behavior of these
distributions in systems that are simple enough to permit both a
theoretical analysis and a precise numerical simulation.

The time of  first return of sets (\ref{deftc}) arises in several
circumstances. Since it controls the shortest return time of
points in the set, it plays a crucial role to establish the
asymptotic (exponential) distribution of the return times of all
points to the set $A$, when the measure of the set $A$ goes to
zero, a different and much investigated topic \cite{GS, A1, A2,
A3, HSV, HV, HLV}. In addition, it has been used to define the
{recurrence dimension}, being used as the gauge set function to
construct a suitable Carath\'eodory measure \cite{AFRA, PSV,
AFRA2}.  Finally, it has been related to the algorithmic
information content \cite{BGI}.

Returns of sets is also relevant in applications, like those of
biological interest. In fact, when the space $X$ consists of
sequences of symbols from a finite alphabet (think {\em e.g.} of
DNA sequencing) particular {\em words}, or {\em motifs} have been
found to be related to biological mechanisms like transcription
sites or protein interaction (see for instance \cite{RRS}). It is
then important to quantify the statistical properties of these
words within the genome. In particular, a typical word of length
$n$ (that is, a finite sequence of $n$ letters) will recur within
a time of the order $e^{n h}$, for large $n$, where $h$ is the
metric entropy of the system, as predicted by the Ornstein-Weiss
theorem \cite{OW}. Yet, if we look at the delay of the first
recurrence of the same word, when observed in the whole (possibly
infinite) sequence, this scales as $n$ \cite{STV,ACS}. This time
of first return is precisely the quantity studied in this paper.

The plan of the paper is the following: in the next section we
review a few results that are useful for the understanding of the
paper. For this reason we neither need nor claim completeness. In
Sect. \ref{seccyl} we consider the case of return times in
cylinders for Bernoulli systems. Quite evidently, this is the
simplest setting where to study return times of sets. We first
present a heuristic explanation of the results rigorously proven
in \cite{AV} and \cite{HV2} that permit to compute, via return
times, the R\'enyi entropies. Then, we refine this analysis to
obtain new results on the type of convergence of the conventional
quantities and we introduce a new one, for which convergence is
much faster. Moreover, our theory allows us to obtain a
description of the different asymptotics of $p(n,k)$ in the
$(n,k)$ plane. In Sect. \ref{secexpa} we leave the symbolic
description to enter a geometric setting by considering expanding
maps of the interval. We show how results proven in the symbolic
setting can be adapted to describe the distribution function
$p(\veps,k)$. In particular, we obtain a formula for the
asymptotic behavior of this function when $k$ and $-\log \veps$
grow while keeping a constant ratio, that involves the Lyapunov
exponent and the R\'enyi entropies. While the treatment of Sect.
\ref{secexpa} is tailored on the specific system under
investigation, the following Sect. \ref{secexpa2} introduces a
more general approach, that confirms the results of the previous
section. Following these ideas, in Sect. \ref{seccat} we consider
the case of linear automorphisms of the two-dimensional torus,
that can be investigated completely. We also conjecture a general
form for the constant $k$ over $-\log \veps$ asymptotics described
above. In the Conclusions we review the new results presented in
this paper.

\section{Review of known results and definitions}
\label{secrev}
 Many facts concerning return times of sets are known. In
this section we review a few of these results that are relevant
for the understanding of this paper. If the dynamical system $
(X,\mu,T) $ has positive metric entropy, $ h_{\mu} $, it has been
proven \cite{STV, ACS} that:
\begin{equation}\label{lmtc}
\liminf_{n\rightarrow \infty}\; \frac{\tau({C_n(x)})}{n} \geq 1\;.
\end{equation}

The limit of the quantity above exists and is equal to one
$\mu$-almost everywhere in certain cases, including irreducible
and aperiodic subshifts of finite type, systems verifying the
specification property \cite{ACS} and even non-uniformly
hyperbolic maps of the interval \cite{HSV,FHV}. It is therefore of
importance to study the measure of the set of points that deviate
from the almost-sure behavior, a quantity that typically decays
exponentially in time. To do this precisely, one defines the
deviation function $M(\delta)$,
\begin{equation}\label{dev1}
M(\delta) :=  \lim_{n\rightarrow \infty}\;
 \frac{1}{n} \log
  \mu(\{ x \in X \; \mbox{ s.t } \; \frac{\tau({C_n(x)})}{n}  < \delta  \}).
\end{equation}
In the language of the previous section this becomes
\begin{equation}\label{smi1}
  M(\delta) = \lim_{n \rightarrow \infty} \frac{1}{n} \log P(n,\delta n)
\end{equation}

In \cite{AV} it has been proven that, for $\psi$-mixing dynamical
systems with some restrictions (see the original paper for definition and further specifications)
the limit above exists and is related to the generalized R\'enyi entropies
$H$ of the invariant measure $\mu$ via
\begin{equation}\label{dev2}
M(\delta)  = (\delta - 1) H( \frac{1}{\delta}-1),
\end{equation}
whenever the latter function exists, being it defined via a summation over
all cylinders $C$ of length $n$,
\begin{equation}\label{dev3}
H(\beta) := - \frac{1}{\beta} \lim_{n\rightarrow \infty}\;
        \frac{1}{n} \log \sum_{C \in \mathcal{C}^n} \mu(C)^{\beta+1}.
\end{equation}
Observe that meaningful values of $\delta$ in eq. (\ref{dev1})
range from zero to one, so that $\beta=\frac{1}{\delta}-1$ is
always larger than zero. For non-integer values of
$\frac{1}{\delta}$, a linear interpolation of the values provided
by eq. (\ref{dev2}) applies.

R\'enyi entropies have been introduced in \cite{RE} and they have
been extensively studied for their connections with various
generalized spectra of dimensions of invariant sets, see for
instance \cite{GP, BS, ER, BPTV, GP2, CS, PPV, TV, TV2}. The
restrictions in \cite{AV} have been removed in \cite{HV2} and in
this last  paper the R\'enyi entropies have been proved to exist
for a weaker class of $\psi$-mixing measures.

 On the other hand, one might try to answer similar questions when
balls are considered in place of cylinders. For instance, if
$B_\varepsilon(x)$ is the ball of radius $\varepsilon$ centered at $x \in
X$, the natural generalization of the limit (\ref{lmtc}) is the quantity
\begin{equation}\label{lmtb}
\eta(x) := \lim_{\varepsilon \rightarrow
0^+}\;\frac{\tau({B_\varepsilon(x)})}{-\log \varepsilon}.
\end{equation}
For a large class of maps of the interval, it has been proved in
\cite{STV} that $\eta(x)$ exists for $\mu$-almost all $x$ and is
equal to the inverse of $\lambda$, the Lyapunov exponent of the
measure $\mu$. For hyperbolic smooth diffeomorphisms of a compact
manifold a similar result holds, to the extent that
\begin{equation}
\frac{1}{\Lambda^u}\leq\liminf_{\veps \rightarrow
0}\frac{\tau({B_\veps(x)})}{-\log \veps}\leq
\limsup_{\veps\rightarrow0}\frac{\tau({B_\veps(x)})}{-\log
\veps}\leq\frac{1}{\lambda^u},
 \end{equation}
where $\Lambda^u$ and $\Lambda^s$  are the largest  and the smallest
Lyapunov exponents, while $\lambda^u$  is the smallest {\em positive}
Lyapunov exponent and $\lambda^s$ is the largest {\em negative} Lyapunov
exponent, see \cite{STV2}. In the case of diffeomorphisms in two dimensions, the above
formula leads to the equality
\begin{equation}\label{duedim}
\lim_{\veps \rightarrow 0}\frac{\tau({B_\veps(x)})}{-\log
\veps}=\frac{1}{\Lambda^u}-\frac{1}{\Lambda^s}=
\frac{1}{\lambda^u}-\frac{1}{\lambda^s} = \frac{D_1(\mu)}{h_\mu}.
\end{equation}
The last equality above is Young's formula, in which $D_1(\mu)$ is the
information dimension and $h_\mu$ is the metric entropy of the measure
$\mu$. One of the goals of this paper is to generalize the results
(\ref{smi1}), (\ref{dev2}) in the case of balls.

\section{Return times in Bernoulli shifts}
\label{seccyl}

In this section we take a close look at Bernoulli processes, that
are the simplest, yet significant, example of $\psi$-mixing
systems. For these, eq. (\ref{dev2}) holds and the function $H$
can be easily computed. In our view, this result is particularly
significant, because it relates a thermodynamical quantity, the
spectrum of R\'enyi entropies, to the statistics of return times.
Our aim is to investigate the kind of convergence holding for eq.
(\ref{smi1}) and more generally the form of the distribution
$p(n,k)$. We shall also introduce a slightly different quantity
than $P(n,\delta n)$, still based on return times, that also
yields $M(\delta)$ in the limit, but for which convergence is much
faster.

Let us start from notations: we consider the full shift on the
space of sequences of $M$ symbols, $\Sigma
:=\{0,\ldots,M-1\}^{Z_+}$. Let us stipulate that the cylinders in
$\mathcal{C}^n$ can be written as
$C_{\sigma_{0},\ldots,\sigma_{n-1}}$, where $\sigma_i \in
\{0,\ldots,M-1\}$, for $i=1,\ldots,n-1$. For short, we shall
sometimes write $\sigma$ for the vector of indices $\sigma_i$, and
$|\sigma|$ will be the length of this vector. $\sigma$ is also
called a ``word'' in symbolic language. With a slight abuse of
notation, we shall sometimes write $\tau(\sigma)$ and
$\mu(\sigma)$, the return time and the measure of the word
$\sigma$, in place of $\tau(C_\sigma)$ and $\mu(C_\sigma)$, the
return time and the measure of the cylinder $C_\sigma$ labeled by
the word $\sigma$. Similarly, we shall let $\Sigma_n
:=\{0,\ldots,M-1\}^{n}$ indicate the set of words of length $n$
and ${\mathcal C}^n$ the set of the associated cylinders, at times
interchangeably.

The Bernoulli invariant measure on $\Sigma$ is induced by the set
of probabilities $\{ \pi_j\}$, $j=0,\ldots,M-1$. For simplicity,
in the numerical simulations, we shall consider the two-symbols
(coin toss) Bernoulli game of parameter $q$: $\pi_0=q, \pi_1=1-q$.

\begin{figure}[ht]
\includegraphics[width=8cm,height=12cm,angle=270]{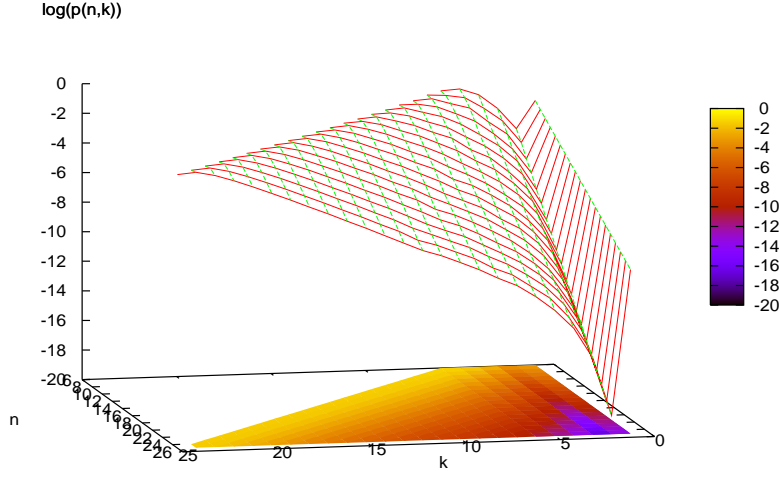}
\caption{Distribution function $p(n,k)$ for the Bernoulli game
with $q=0.3$.} \label{figsymb1}
\end{figure}

Fig. \ref{figsymb1} depicts the function $p(n,k)$ for $q=0.3$. Two
features are evident: the first, is the slow decay of $p(n,1)$
with increasing $n$. The second is the much faster decay of
$p(n,2)$. Both these features, and more, can be explained by
computing the asymptotic behavior of $p(n,k)$ for $k$ fixed and
large $n$.

The case $k=1$ follows immediately from the observation that the only
words $\sigma$ for which $\tau(\sigma)=1$ are composed of a single symbol:
$\sigma_i=j$ for all $i$, where $j \in \{0,\ldots,M-1\}$. The measure of
the associated cylinders is $\pi_j^n$, so that $p(n,1)$ is the sum of
these quantities over all $j$, and behaves asymptotically as $a^n$, where
$a = \max \{ \pi_j \}$.

Let now ${\mathcal T}_{n,k}$ the set of words of length $n$ with first
return time $k$:
\begin{equation}\label{idea00}
 {\mathcal T}_{n,k} = \{ \sigma
 \in \Sigma_n \mbox{ s.t. }
  \; \tau(\sigma) = k \}.
\end{equation}
The first observation is that for any $\sigma \in {\mathcal T}_{n,k}$ the
symbols $\sigma_i$ repeat periodically with period $k$:
\begin{equation}\label{idea0}
   \sigma_{k+j} = \sigma_j, \;\; j=0, \ldots, n-k .
\end{equation}
Define the periodic replication operator $P^{n,k}$ as that which takes any
word $\sigma$ of length $k$ into the word $\sigma' = P^{n,k}(\sigma)$ of
length $n \geq k$, that satisfies eq. (\ref{idea0}). It is of relevance
now to consider the set of words ${\mathcal W}_{n,k}$ defined implicitly
by
\begin{equation}\label{idea01}
 {\mathcal T}_{n,k} = P^{n,k} ({\mathcal W}_{n,k}).
\end{equation}
The meaning of this definition is simple: words in ${\mathcal W}_{n,k}$
have length $k$ and are all and the only periodic ``roots'' of words of
${\mathcal T}_{n,k}$:
\begin{equation}\label{idea02}
 {\mathcal W}_{n,k} = \{ \sigma \in \Sigma_k \mbox{ s.t. }
   \; \tau(
 P^{n,k} (\sigma)) = k \}.
\end{equation}
Controlling ${\mathcal W}_{n,k}$ is then the same thing as controlling
${\mathcal T}_{n,k}$.

It is not difficult to prove that for any $k$
\begin{equation}\label{idea03}
 {\mathcal T}_{k,k}  = {\mathcal W}_{k,k}   \subset
{\mathcal W}_{k+1,k}   \subset
 \ldots
 {\mathcal W}_{2k,k}   =
 {\mathcal W}_{2k+1,k}   = \ldots =
 {\mathcal W}_{2k+j,k},
\end{equation}
for any $j \geq 0$. Therefore, when $n \geq 2k$ the set of ``roots'' $
{\mathcal W}_{n,k}$ is constant in $n$. Define
\begin{equation}\label{idea04}
  m(k) = \max_{\sigma \in {\mathcal W}_{2k,k}}
   \{ \# \{ j \mbox{ s.t. } \sigma_j = 0 \} \}.
\end{equation}
This is the maximum number of zeros in a word in ${\mathcal W}_{2k,k}$.
Since all sets ${\mathcal W}_{n,k}$ are invariant for permutation of the
symbols $\{0,\ldots,M-1\}$, $m(k)$ is also the maximum number of any other
symbol in a word in ${\mathcal W}_{2k,k}$. Suppose now that only one
symbol in $\{0,\ldots,M-1\}$ has probability $a$, so that the probability
of all other symbols is less than $a$. For simplicity, let us only
consider the case of $M=2$. Then, the probability of the word for which
$m(k)$ is obtained gives the leading term in the asymptotic of $p(n,k)$:
\begin{equation}\label{idea05}
\log(p(n,k)) \sim \frac{n}{k} [{m(k)} \log a + (k-m(k)) \log(1-a)].
\end{equation}
It is also rather easy to see that $m(k)=k-1$. In fact, $m(k)=k$ is
possible only for $k=1$, and $\sigma_{k-1}=1$, $\sigma_j=0$ for
$j=0,\ldots,k-2$ is a word in ${\mathcal W}_{2k,k}$. Then,
\begin{equation}\label{idea06}
\frac{1}{n}  \log(p(n,k)) \sim (1-\frac{1}{k}) \log a + \frac{1}{k}
\log(1-a).
\end{equation}
This formula gives the decay rates of $p(n,k)$. They are increasing
functions of $k \geq 2$, the most negative being precisely that of
$p(n,2)$, and tend to $\log(a)$ when $k$ goes to infinity.

Let us now consider the asymptotic behavior of $p(n,k)$ over the
line $k=\delta n$, where $0< \delta < 1$. Firstly, it is clear
that ${\mathcal W}_{n,k} \neq \Sigma_k$, since some words of
length $k$ may be associated with {\em shorter} return times than
$k$, and one the technical achievements of refs. \cite{AV, HV2} is
to deal properly with this issue. This fact notwithstanding, we
may begin by assuming heuristically that among all words $\sigma$
in $\Sigma_k$, those associated with return times smaller than $k$
and therefore not in ${\mathcal W}_{n,k}$, are statistically
negligible, in some limit. More refined arguments will follow
later in this section. Under this assumption, the probability
$p(n,k)$ can be approximated by a sum over all words of length
$k$:
\begin{equation}\label{idea1}
  p(n,k) =
    \sum_{\sigma \in {\mathcal W}_{n,k}} \mu
  (P^{n,k}(\sigma)) \simeq
  \sum_{\sigma \in \Sigma_k }  \mu
  (P^{n,k}(\sigma)).
\end{equation}
Formula (\ref{idea1}) can be further developed by estimating the cylinder
measure $\mu (\sigma')$, where $\sigma'=P^{n,k}(\sigma)$. Since
$|\sigma|=k$, $|\sigma'|=n$, and since $\sigma'$ satisfies eq.
(\ref{idea0}),
\begin{equation}\label{idea2}
   \log \mu ({\sigma'}) = \frac{n}{k} \log \mu (\sigma),
\end{equation}
exactly whenever $\frac{n}{k}$ is an integer (and approximately in the
other cases) so that
\begin{equation}\label{idea3}
  p(n,k) = \sum_{\sigma \in \Sigma_k} \mu
  (C_\sigma)^{n/k}.
\end{equation}
If we now set $k=\delta n$, with $0 < \delta \leq 1$ and $\delta$ the
inverse of an integer, we can estimate
\begin{equation}\label{idea4}
  \lim_{n \rightarrow \infty} \frac{1}{n} \log p(n,\delta n) =
  \delta \lim_{k \rightarrow \infty} \frac{1}{k}
  \log (\sum_{\sigma \in \Sigma_k} \mu
  (C_\sigma)^{1/\delta}).
\end{equation}
If we now compare this equation with the definition of R\'enyi
entropies, eq. (\ref{dev3}), we easily obtain that
\begin{equation}\label{idea6}
  M(\delta) = \lim_{n \rightarrow \infty} \frac{1}{n} \log p(n,\delta n) =
  (\delta-1) H (\frac{1}{\delta} - 1).
\end{equation}

\begin{figure}[ht]
\includegraphics[width=8cm,height=12cm,angle=270]{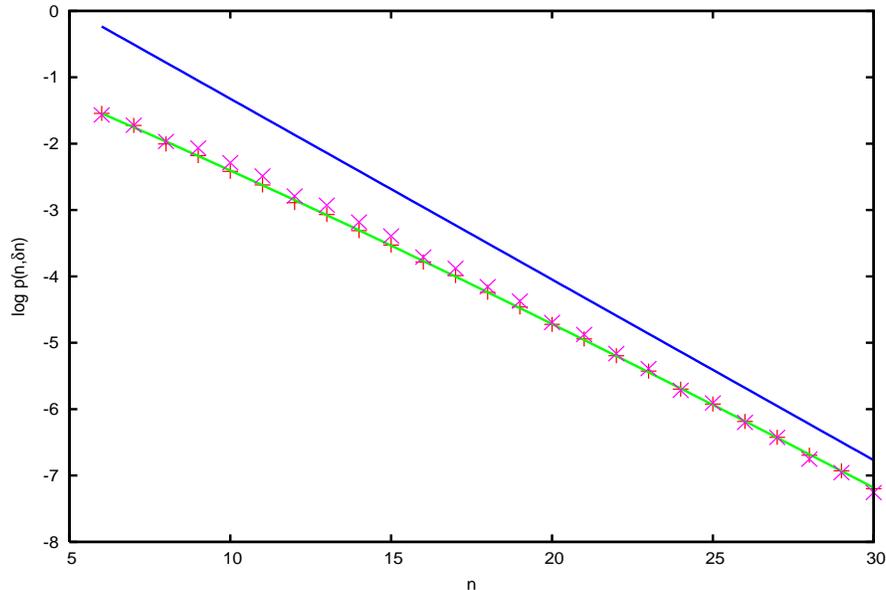}
\caption{Distribution function $p(n,\delta n)$ for a Bernoulli
process with $q=0.3$, $\delta=0.5$ (X) and cumulative distribution
function $P(n,\delta n)$ of the same process (crosses). The first
function has been shifted upwards by a fixed quantity to match the
latter. Both sets of data are consistent with the behavior
described in the text: The straight blue line has indeed slope
$(\delta-1) H (\frac{1}{\delta} - 1)$, and the green fitting curve
(which becomes asymptotically tangent to the blue line) is given
by eq. (\ref{idea6b}).} \label{figr6-dp5}
\end{figure}

Observe that this heuristic result has been derived for $p(n,k)$
rather than $P(n,k)$, for which it is known to hold
rigorously--modulo the linear interpolation required for
non-integer values of $1/\delta$. Indeed, numerical experiments on
Bernoulli schemes with two symbols (coin toss) show that formula
(\ref{idea6}) holds. For instance, Figure \ref{figr6-dp5} shows
the logarithm of both $p(n,\delta n)$ and $P(n,\delta n)$ versus
$n$, for $q=0.3$ and $\delta=0.5$, compared with a straight line
$g_\delta(n) = an +b$ with slope $a=(\delta-1) H (\frac{1}{\delta}
- 1)$. Both quantities are well fitted by a function of the kind
$g_\delta(n) + c e^{d n}$.

\begin{figure}[ht]
\includegraphics[width=8cm,height=12cm,angle=270]{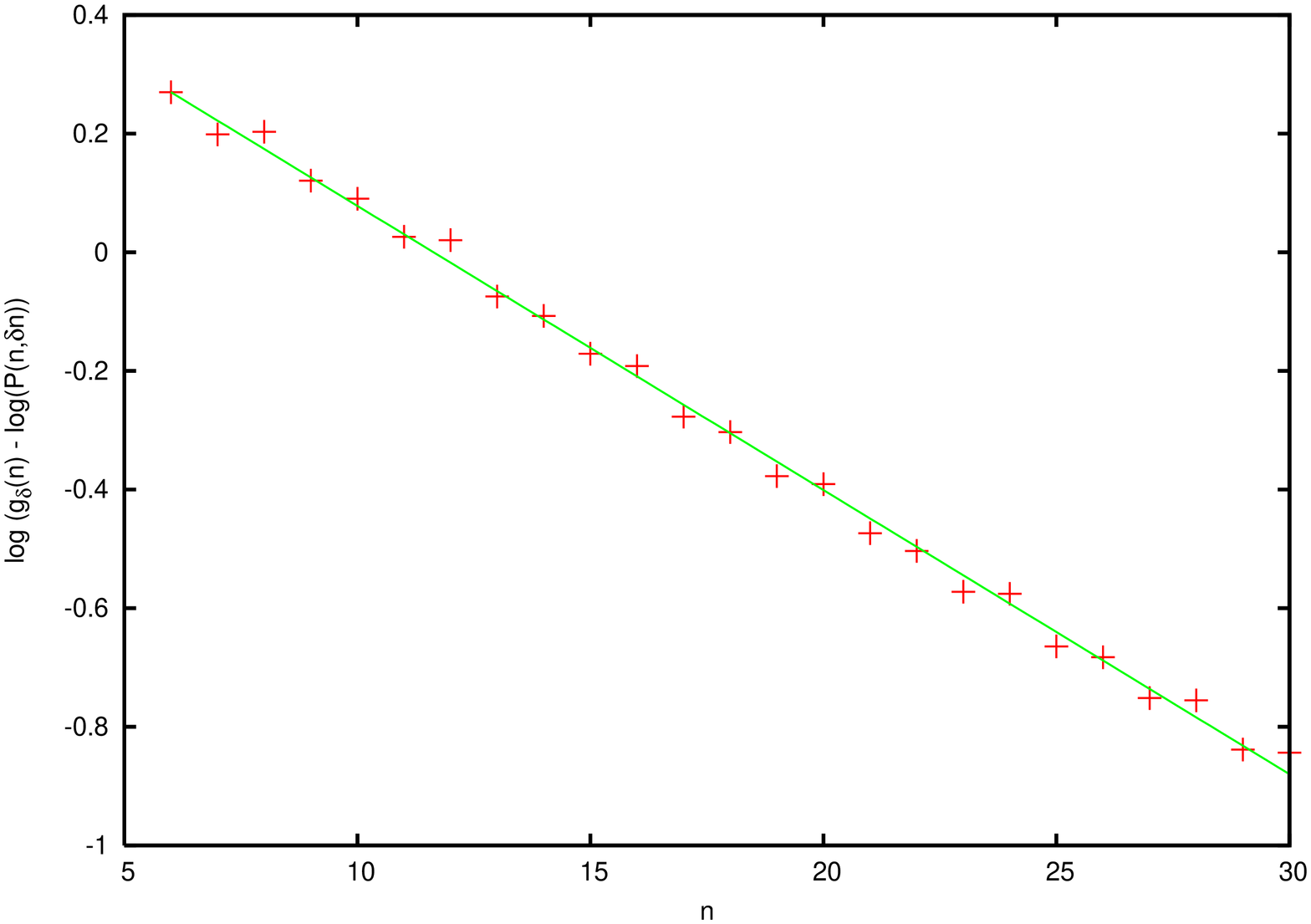}
\caption{Logarithm of the difference between the logarithm of the
cumulative distribution function $P(n,\delta n)$ of the Bernoulli
process with $q=0.3$, $\delta=0.5$ and the straight line
$g_\delta(n)$ in Fig. \ref{figr6-dp5}. The fitting  straight
(green) line implies the exponential decay used in the fit in Fig.
\ref{figr6-dp5}.} \label{figr6-dp5-dif}
\end{figure}

To validate this picture, confirmed by numerical experiment on other
values of $q$ and $\delta$ (for $\delta$ not equal to the inverse of an
integer, a linear interpolation formula applies \cite{AV, HV2}) we plot in
Fig. \ref{figr6-dp5-dif} the logarithm of the difference between data and
the straight line in Fig. \ref{figr6-dp5}. An exponential decay is clearly
observed.

In conclusion, numerical experimentation support the hypothesis that the
asymptotic behavior in eq. (\ref{idea6}) is attained with a decaying
exponential term parameterized by the constants $c$ and $d<0$, together
with a slowly decaying contribution arising from the constant $b$:
\begin{equation}\label{idea6b}
  \frac{1}{n} \log p(n,\delta n) \simeq M(\delta) + \frac{b}{n} + \frac{c}{n} e^{dn}.
\end{equation}
We shall introduce momentarily a new quantity to improve on this kind of
convergence.

The same heuristic arguments imply an approximate form for the behavior of
the distribution function for different $n$ and $k$. Carrying out the
computations of eq. (\ref{idea3}) together with eq. (\ref{dev3}) we obtain
\begin{equation}\label{idea9}
  \log (p(n,k))  \sim (k-n) H(\frac{n}{k}-1).
\end{equation}

\begin{figure}[ht]
\includegraphics[width=8cm,height=12cm,angle=270]{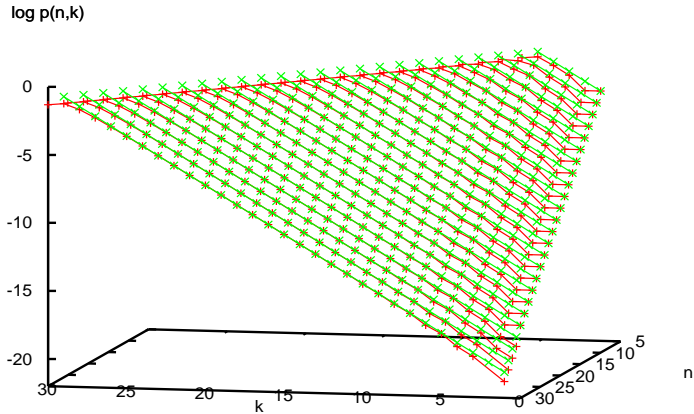}
\caption{Distribution function $p(n,k)$ (crosses, red lines) and the
approximation function in eq. (\ref{idea9}) (x, green line) versus $n$ and
$k$ in the case of a Bernoulli game with $q=1/2$.} \label{figr7-3a}
\end{figure}

We can test this approximation in the case of the Bernoulli scheme with
$q={1/2}$. Here, trivially the approximation (\ref{idea9}) becomes $\log
(p(n,k))  \sim (k-n) \log (2))$. This function fits almost perfectly the
numerical data in Fig. \ref{figr7-3a}.

\begin{figure}[ht]
\includegraphics[width=8cm,height=12cm,angle=270]{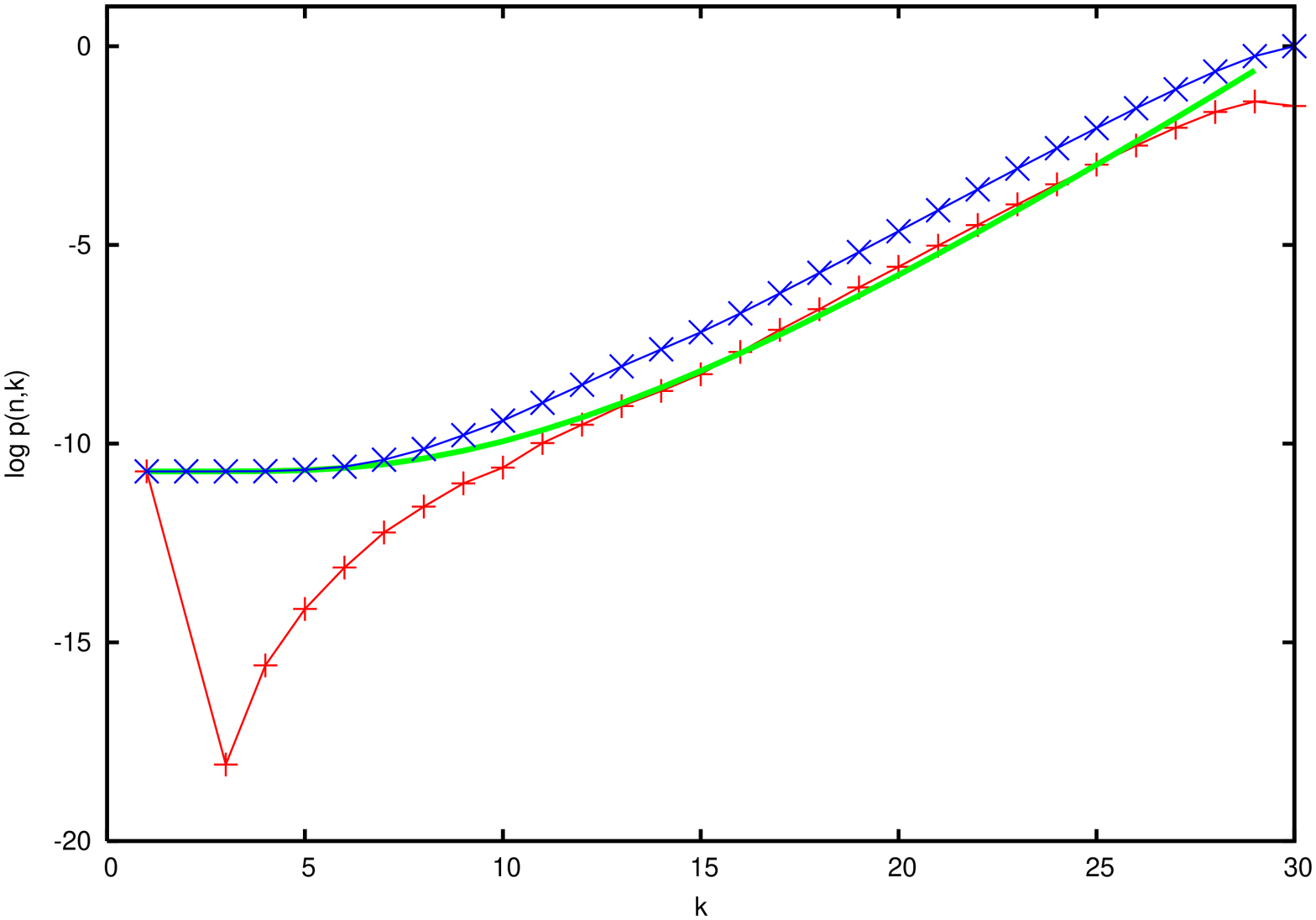}
\caption{Distributions $P(n,k)$ (X, blue line), $p(n,k)$ (crosses,
red line) and the approximation function in eq. (\ref{idea9})
(green line) versus $k$ for $n=30$, in the case of a Bernoulli
process with $q=0.3$} \label{figr6-3a}
\end{figure}

A less favorable case is offered by the Bernoulli scheme with $q={.3}$. In
Fig. \ref{figr6-3a} we plot $P(n,k)$ and $p(n,k)$  versus $k$ for $n=30$,
together with the approximation provided by eq. (\ref{idea9}) We notice
that this latter fits well both functions at $k=1$ (quite obviously, being
this behavior associated with the measure of the cylinder of unity return
time, see above), while approximation stays reasonable only for $P(n,k)$
at small values of $k>1$. Then, in the intermediate region the slope of
both curves agrees with the interpolation while the numerical value only
for $p(n,k)$.

Certainly, the agreement observed in the last figure is far from
satisfactory. The reason is to be found in the approximation made in eq.
(\ref{idea1}).
%, rather than in the more reliable estimate in eq. (\ref{idea2}).
The same fact is at the origin of the slow convergence observed in eq.
(\ref{idea6b}).
We conclude this section unveiling these reasons and providing a more
rigorous and insightful treatment of the problem. This improvement is
inspired by the idea of summation over prime periodic orbits in dynamical
systems.

As we mentioned, the approximation in eq. (\ref{idea1}) above is
based on the idea that the roles of $\Sigma_k$ and ${\mathcal
W}_{n,k}$ can be interchanged, the effects of their difference
being negligible in the limit. The price to pay in this procedure
is the slowly decaying term in the asymptotics (\ref{idea6b}). We
can be more careful: indeed, we can show that $\Sigma_k$ can be
rigorously partitioned into the periodic repetition of {\em
different} sets ${\mathcal W}_{n,k'}$. The lemma is the following:
for any $n \geq 2k$
\begin{equation}\label{idea10}
\Sigma_k = \bigcup_{k'|k} P^{k,k'}({\mathcal W}_{n,k'}),
\end{equation}
where the union is over all integer $k'$ that divide $k$ and where the
sets $P^{k,k'}({\mathcal W}_{n,k'})$ are the full completion of
$\frac{k}{k'}$ cycles of the word of length $k'$. These sets are pairwise
disjoint.

On the basis of this lemma, we can reverse the ordering in eq.
(\ref{idea1}) to get:
\begin{equation}\label{idea12c}
\sum_{\sigma \in \Sigma_k }  \mu
  (P^{n,k}(\sigma)) =
  \sum_{k'|k} \sum_{\sigma \in {\mathcal W}_{n,k'}} \mu
  (P^{n,k} \circ P^{k,k'} (\sigma)).
\end{equation}
Since $k'$ divides $k$, $P^{n,k} \circ P^{k,k'} = P^{n,k'}$ and
therefore the chain of equalities continues with
\begin{equation}\label{idea12b}
  \sum_{\sigma \in \Sigma_k }  \mu
  (P^{n,k}(\sigma)) = \sum_{k'|k} \sum_{\sigma \in {\mathcal W}_{n,k'}} \mu
  (P^{n,k'} (\sigma)) =
 \sum_{k'|k} \sum_{\sigma \in {\mathcal T}_{n,k'}} \mu
  (\sigma),
\end{equation}
where we have used eq. (\ref{idea01}). Finally,
\begin{equation}\label{idea13}
\sum_{\sigma \in \Sigma_k }  \mu
  (P^{n,k}(\sigma)) =
  \sum_{k'|k} \sum_{\sigma \in {\mathcal T}_{n,k'}} \mu
   (\sigma) = \sum_{k'|k} p(n,k').
\end{equation}
On the other hand,
\begin{equation}\label{idea14d}
% \sum_{\sigma \in \Sigma_k }
 \mu (P^{n,k}(\sigma)) =
 % \sum_{\sigma \in \Sigma_k }
    (\mu(C_\sigma))^\frac{n}{k},
\end{equation}
exactly when $\frac{n}{k}$ is an integer, and approximately otherwise, so
that choosing $k/n=\delta$, so that $\delta n = k$ is an integer, and
defining the new quantity
\begin{equation}\label{idea15a}
  Z_p(\delta,n) :=
    \sum_{k'| \delta n} p(n,k'),
\end{equation}
we find that
\begin{equation}\label{idea15s2}
  M_p(\delta) = \lim_{n \rightarrow \infty} \frac{1}{n} \log (
    Z_p(\delta,n)) =
  (\delta-1) H (\frac{1}{\delta} - 1).
\end{equation}
This formula is valid for all $0<\delta\leq \frac{1}{2}$ that are the
inverse of an integer. For the other values, the linear interpolation
between the values at the nearest inverses of an integer applies.

\begin{figure}[ht]
\includegraphics[width=8cm,height=12cm,angle=270]{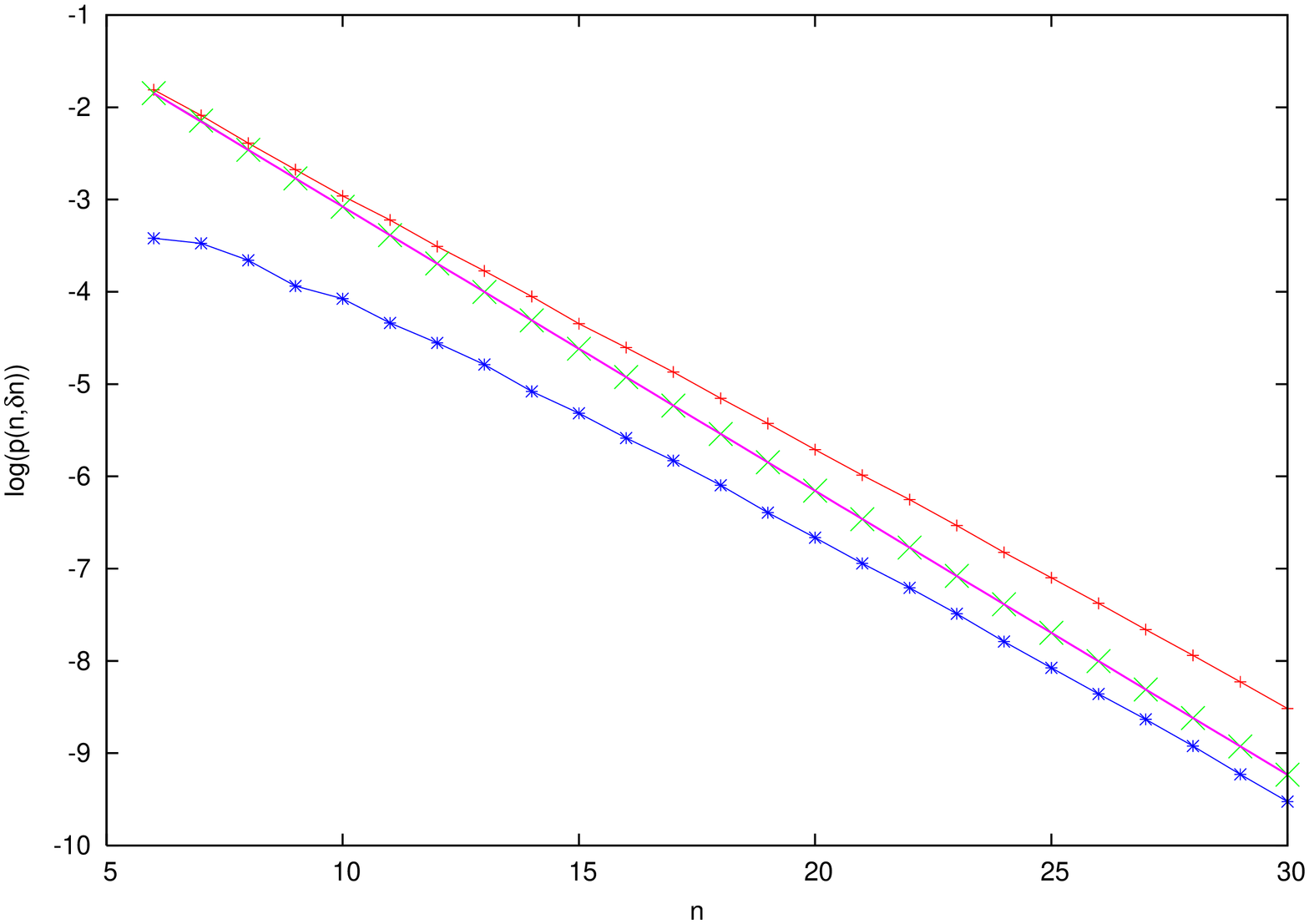}
\caption{Logarithm of the cumulative distribution function
$P(n,\delta n)$ (red line, pluses) and of the distribution
function $p(n,\delta n)$ (blue line, stars) together with $\log
Z_p(\delta,n)$ (large crosses, light blue) and the line
$g_\delta(n)=M(\delta) n$ (magenta) for the Bernoulli process with
$q=0.3$, $\delta=0.4$ } \label{figd=p4}
\end{figure}

\begin{figure}[ht]
\includegraphics[width=8cm,height=12cm,angle=270]{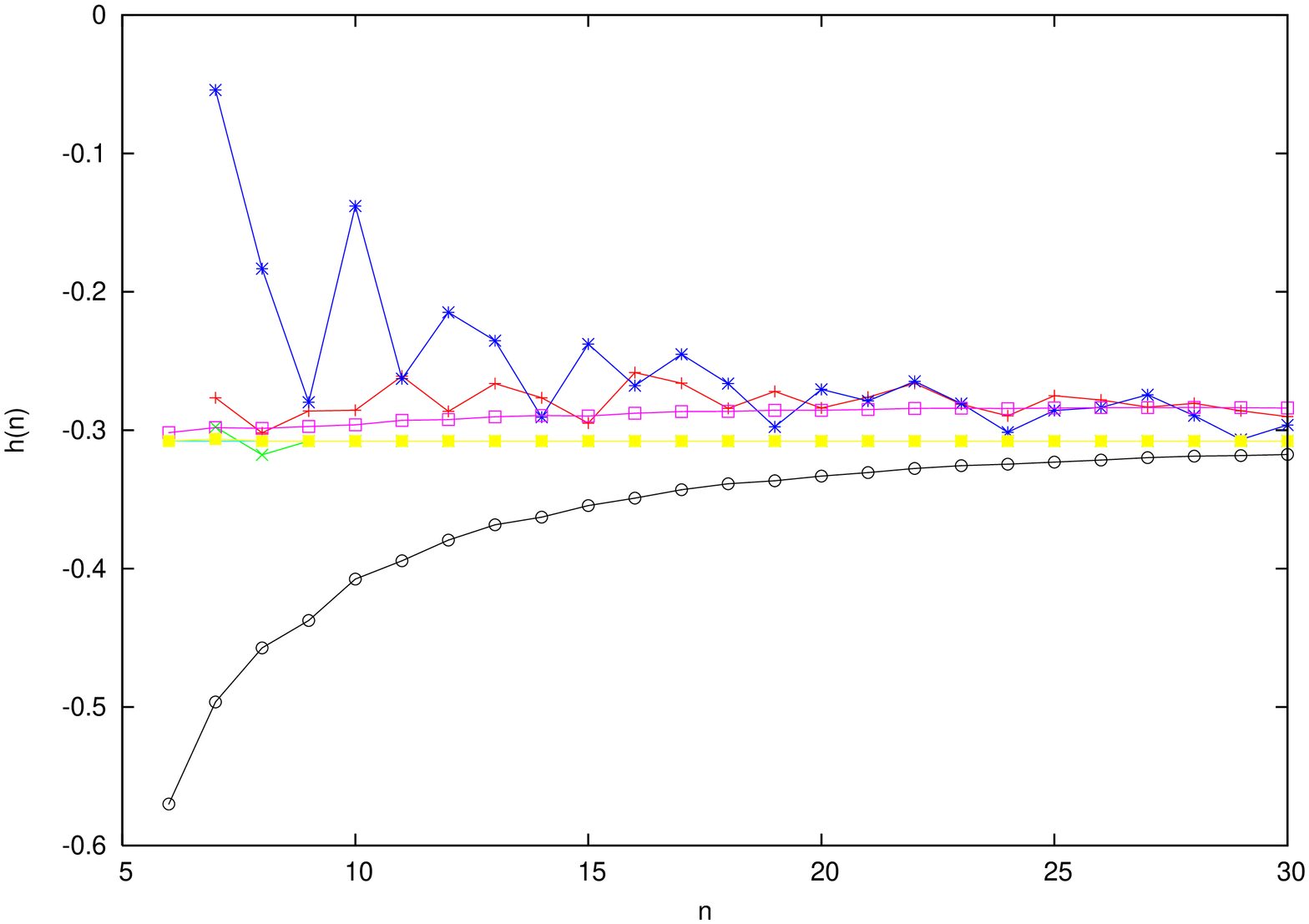}
\caption{Various functions for  of the Bernoulli process with
$q=0.3$, $\delta=0.4$. Start with $\log(p(n,\delta n))/n$: black
curve, circles; $\log(p(n,\delta n))-\log(p(n-1,\delta (n-1))$:
blue curve, stars; $\log(P(n,\delta n))/n$ open squares, magenta;
$\log(P(n,\delta n))-\log(P(n-1,\delta (n-1)))$: crosses, red
curve; $\log(Z_p(n,\delta))/n$: yellow squares;
$\log(Z_p(n,\delta))-\log(Z_p(n-1,\delta))$: crosses, green curve.
The last two set of data sit to numerical precision on the line
$h=-0.307795889757108$ that is obtained by the interpolation
formula. } \label{figd=p4-2}
\end{figure}

Numerical verification follows: in Fig. \ref{figd=p4} we plot the
logarithm of $P(n,\delta n)$, $p(n,\delta n)$ and of the period
summation function $Z_p(\delta,n)$ versus $n$ for the Bernoulli
process with $q=0.3$. The line $g_\delta(n)=M(\delta) n$ fits
almost perfectly the last set of data. On the other hand, the
other two functions display the slow convergence discussed above.
To further appreciate the improvement brought about by using $Z_p$
consider Fig. \ref{figd=p4-2}, where we plot the difference
between successive values of the logarithm of the above functions,
and these logarithms divided by $n$. All these quantities have
limit $M(\delta)$. We compute this value from the linear
interpolation of $(\delta-1) H(\frac{1}{\delta}-1)$ to get the
value tabulated. The distributions $p$ and $P$ converge slowly,
while $Z_p$ gives a numerically exact result. In conclusion, the
term $b>0$ plaguing the convergence in eq. (\ref{idea6b}) was due
to approximate counting and does not show up for the newly
introduced quantity.

\section{Expanding maps of the interval}
\label{secexpa}

In this section, we study a family of one-dimensional dynamical systems
for which we can derive both a formula for the asymptotic distribution and
a generalization of the deviation result. This family will also serve to
begin to understand the dynamical phenomena occurring when considering
{\em ball}, rather than cylinder, return times: one has to match geometry
and dynamics. Further detail will be added in the following section.

We start by constructing a family of measures supported in $[0,1]$
by means of an {\em affine iterated function system}: given a set
of $M$ non-overlapping intervals $I_j=[a_j,b_j] \subset [0,1]$
$j=0,\ldots,M-1$, define the lengths $\delta_j = b_j - a_j$, and
construct the affine maps
\begin{equation}\label{ifs1}
 \phi_j(x) := \delta_j x + a_j, \; j=1,\ldots,M.
\end{equation}
Each map $\phi_j$ takes $[0,1]$ into $[a_j,b_j]$. Consider then
the set action $\Phi$ that maps the set $A \subset [0,1]$ into
$\Phi(A) := \bigcup_{j=0}^{M-1} \phi_j(A)$. Repeated action of
$\Phi$ on $[0,1]$ defines a Cantor set $\mathcal S$ in $[0,1]$:
\begin{equation}\label{ifs2}
 {\mathcal S} := \bigcap_{k=1}^{\infty} \Phi^k([0,1]).
\end{equation}

We can then construct a family of measures whose support is this
set $\mathcal S$: let us choose real numbers
$\{\pi_j\}_{j=1,\ldots,M}$ such that $\pi_j>0$, $\sum_{j=1}^M
\pi_j = 1$, and consider the unique measure $\mu$ for which
\begin{equation}
 \int f(s) d \mu(s) =  \sum_{j=0}^{M-1}
  \pi_j
  \int (f \circ \phi_j) (s) d \mu(s),\label{balabiot}
  \end{equation}
holds for any continuous function $f$. It is then easy to show that, for
any choice of the set of real numbers $\{\pi_j\}$, the measure $\mu$ is
mixing for the piece-wise linear transformation $T$ defined on $S$ by:
\begin{equation}
  T(x) =  \frac{1}{\delta_j} (x - a_j), \mbox { if } x \in
  I_j=[a_j,b_j] .\label{balab2}
  \end{equation}
In fact, the maps $\{\phi_j\}$ turn out to be the inverse branches of $T$.
As such, they can be employed to build the cylinders $\mathcal{C}^n$ of
this dynamical system: letting $\sigma_i \in \{0,\ldots,M-1\}$, for
$i=1,\ldots,n-1$ these latter can be labelled as
 \begin{equation}
  C_{\sigma_{0},\ldots,\sigma_{n-1}} = (\phi_{\sigma_0} \circ \cdots
  \circ \phi_{\sigma_{n-1}} ) ([0,1]).\label{balab3}
  \end{equation}
Following eq. (\ref{ifs1}), the geometric length
$\ell(C_{\sigma})$ is easily computed:
\begin{equation}
   \ell(C_{\sigma_{0},\ldots,\sigma_{n-1}}) =
\prod_{i=0}^{n-1}
  \delta_{\sigma_i}.\label{balab4a}
  \end{equation}
Equally easily, because of eq. (\ref{balabiot}), the measure
$\mu(C_{\sigma})$ is
\begin{equation}
  \mu(C_{\sigma_{0},\ldots,\sigma_{n-1}}) =
 \prod_{i=0}^{n-1}
  \pi_{\sigma_i}.\label{balab4}
  \end{equation}
Therefore, this dynamical system is metrically equivalent to a Bernoulli
shift on $M$ symbols, with probabilities $\pi_i$, $i=1,\ldots,M-1$, that
we have discussed in section \ref{seccyl}. Yet, when examining the
distribution of {\em ball} return times, we must investigate is the
geometrical relation between balls of a fixed radius and cylinders of
different symbolic length.

\begin{figure}[ht]
\includegraphics[width=8cm,height=12cm,angle=270]{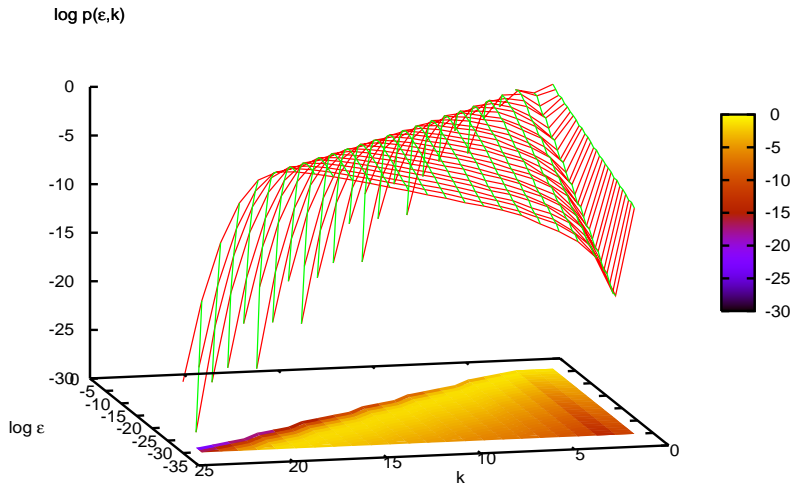}
\caption{Distribution function $p(\varepsilon,k)$ for for the
dynamics of the two map IFS with $\delta_1=0.3$, $\delta_2=0.2$,
$\pi_1=0.3$ and $\pi_2=0.7$. %run5
 }
\label{figr5-2}
\end{figure}

In order to analyze this relation, we observe that any ball
$B_\varepsilon(x)$ can be written as a union over cylinders of an
appropriate (fixed) length $n$: that is, for all $x$ and
$\varepsilon$ there exist $n$ and a collection of indices $\sigma
\in I$, $|\sigma|=n$ such that

\begin{equation}\label{cover1}
B_\varepsilon(x) = \bigcup_{\sigma \in I} C_\sigma.
\end{equation}

This is a consequence of the fact that these measures are singular
w.r.t. Lebesgue and their support have gaps of positive length.
Therefore, for sufficiently large $n$, the two boundary points of
the ball end up in the closure of a gap in the support of $\mu$.

Furthermore, since both $n$ and $I$ depend on the ball under
consideration, we define the {\em symbolic length} of
$B_\varepsilon(x)$ as
\begin{equation}\label{cover2}
N_\varepsilon(x) = n - \max \{ j \in {\bf Z} \mbox{ s.t. } M^j
\leq \# (I) \},
\end{equation}
where $\# (I)$ is the cardinality of the set $I$ and where $M$, as
before, is the number of inverse branches of $T$.  The idea behind
this definition is to measure a sort of effective length of the
cover of $B_\varepsilon(x)$. For instance, if eq. (\ref{cover1})
would require $\# (I) = 4$ cylinders of length $n=3$ with $M=2$ we
would effectively consider this union as if it were a single
cylinder of length $N_\veps=1$.

\begin{figure}[ht]
\includegraphics[width=8cm,height=12cm,angle=270]{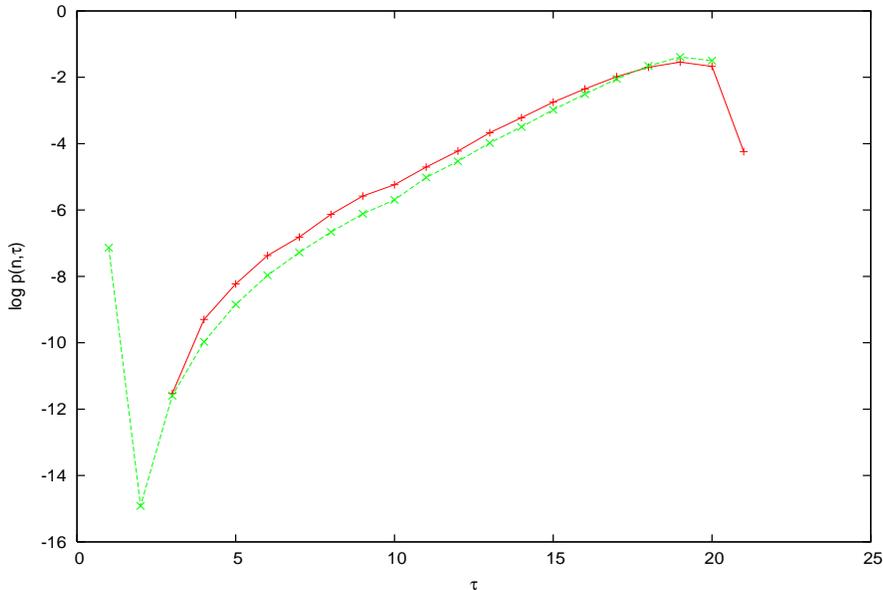}
\caption{Distribution function $p(n,\tau)$ for return times of
balls of radius $\varepsilon = 8.989 \; 10^{-12}$ and
$n=N_\varepsilon=20$ (red line, crosses), for the dynamics of the
two map IFS described in Fig. \ref{figr5-2}. It is compared with
the distribution function $p(n,\tau)$, $n=20$ of the Bernoulli
process with $q=0.3$ (green line, X's). See text for details.}
\label{figr5-1}
% analogous figure r5-1.ps for run5 data
%% analogous figure r3-2.ps for run5 data
%
\end{figure}

This definition is instrumental in formulating a working hypothesis: we
surmise that the statistical distribution of return times of boxes
$B_\varepsilon(x)$ characterized by the same symbolic length
$N_\varepsilon(x)=n$ will scale as that of cylinders (again, of that given
length $n$) in a Bernoulli shift. This hypothesis is confirmed by
numerical computation. At a fixed radius $\veps$, we compute the return
time distribution for {\em all} points $x$ with a fixed symbolic length,
computed via eq. (\ref{cover2}) and we compare it with the discrete
distribution of the corresponding symbolic Bernoulli process. Data
reported in Figure \ref{figr5-1} are obtained for a two map I.F.S.
dynamics. Accordance is significant.

Therefore, to obtain the distribution of return times of balls of
fixed radius $\veps$ we must know the cylinder return time
distribution, $p(n,k)$, discussed in Sect \ref{seccyl}, but also
the measure of center points whose balls have a given symbolic
length:
  \begin{equation}\label{cover3}
   \Psi(\veps,n) = \mu ( \{ x \mbox{ s.t. }
N_\varepsilon(x) = n \}).
  \end{equation}
In fact, from this information, we obtain
\begin{equation}\label{cover8}
   p(\veps,k) =
   \sum_{n} \Psi(\veps,n) p(n,k).
  \end{equation}

To estimate $\Psi(\veps,n)$, we consider, this time at fixed $n$,
the distribution of the geometric lengths of cylinders,
$\ell(C_\sigma)$: let $z=\log(\veps)$, and define
  \begin{equation}\label{cover5}
   \psi(z,n) := \frac{d}{dz} \mu ( \{ x \mbox{ s.t. }
   \log(\ell(C_n(x))) \leq z \} ).
  \end{equation}
When $n$ is sufficiently large, this can be approximated by a
continuous distribution, precisely, by a normal distribution
$\mathcal{N}_{-\lambda n,S \sqrt{n}}(z)$ of mean $-\lambda n$, and
variance $S^2 n$, where $\lambda$ is the Lyapunov exponent of
$\mu$, and $S$ the standard deviation of the multiplicative
process. Both quantities can be easily computed in this case:
\begin{equation}\label{cover6}
   \lambda = - \sum_{j=0}^{M-1}
  \pi_{j} \log(\delta_j),
  \end{equation}
and
\begin{equation}\label{cover7}
   S^2 =  - \lambda^2 + \sum_{j=0}^{M-1}
  \pi_{j} (\log(\delta_j))^2.
  \end{equation}

We now conjecture that we can exchange the role of $z$ and $n$ in
this derivation, so that $\Psi(\veps,n)$ (with $\veps$ fixed) be
approximated by the distribution $\psi(z,n)$ (with $z=\log(\veps)$
fixed) when properly normalized and, in turn, with
$\mathcal{N}_{-\lambda n,S \sqrt{n}}(z)$, with $z = \log (\veps)$
fixed, and properly normalized via the constant $A$ to yield the
discrete distribution $D_{\veps}(n)$):
  \begin{equation} \label{cover7b}
  D_{\veps}(n) = A \; \mathcal{N}_{-\lambda n,S \sqrt{n}}(\log \veps),
  \;\; \sum_n D_{\veps}(n)=1.
  \end{equation}

\begin{figure}[ht]
\includegraphics[width=8cm,height=12cm,angle=270]{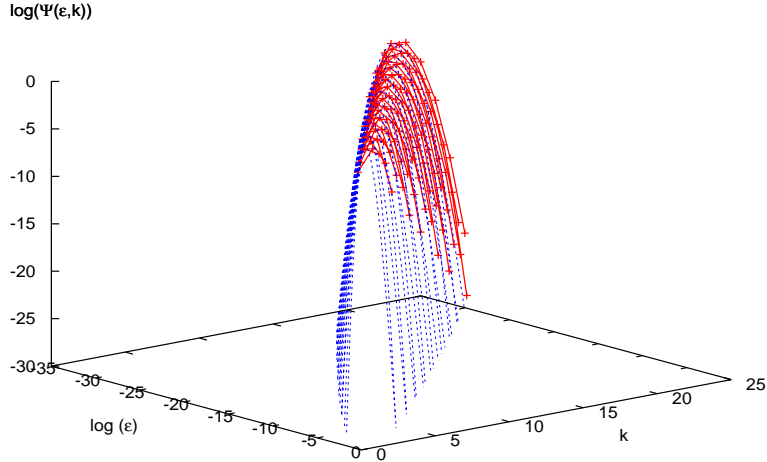}
\caption{Distribution function $\Psi(\veps,n)$ (red lines, points) and
$N(z,n)$ (blue lines) compared (for definitions, see text), in the case of
Fig. \ref{figr5-2}.} \label{r3ps}
% analogous figure f26-r3b.ps for run5 data
%% analogous figure f26-r5.ps for run5 data
%
\end{figure}

This conjecture is validated numerically in Figure \ref{r3ps},
that reports $D_{\veps}(n)$ and $\Psi(\veps,n)$ for the same case
of Figs. \ref{figr5-2}, \ref{figr5-1}. Indeed, for our purposes we
do not need the exact form of the limit distribution, but only its
scaling behavior in $n$ and $\veps$. In conclusion, we can write
the distribution function $p(\veps,k)$ as
\begin{equation}\label{cover10}
   p(\veps,k) = \sum_{n} D_\veps(n) p(n,k),
  \end{equation}
where $D_\veps(n)$ has mean $\bar{n}=-z/\lambda$ and standard
deviation $s=\lambda^{-3/2} S \sqrt{-z} $, sharply localized in
the interval $[n-s,n+s]$. This explains the shape of the graph
reported in Fig. \ref{figr5-2}, that is to be compared with that
of Fig. \ref{figsymb1} in Section \ref{seccyl}: the Gaussian
smoothing is particularly evident near the line
$k=-\log(\veps)/\lambda$.

Finally, eq. (\ref{cover10}) is the basis to derive formulae akin
to those of Sect. \ref{seccyl}. In particular, it validates the
analogue of formula (\ref{idea6}) that becomes the fundamental
result of this section. For one dimensional expanding maps, the
following asymptotic formula holds, that links the asymptotic of
return time distributions to the Lyapunov exponent and to R\'enyi
entropies:
\begin{equation}\label{idea6c}
\lim_{\veps \rightarrow 0} \frac{\log p(\veps,-\delta
\log(\veps)/\lambda)}{\log \veps}
   =
  \frac{1-\delta}{\lambda} H (\frac{1}{\delta} - 1).
\end{equation}
The same observations about the convergence speed of this limit
detailed in Sect. \ref{seccyl} apply here.

\section{A second approach to expanding maps}
\label{secexpa2}

In this section, we present a second, more general approach to the
statistics of first returns in balls for one-dimensional piecewise linear
expanding maps of the type studied in the previous section. This approach
can be fruitfully extended to more general situations and, informal as it
is now, clearly points to the direction where rigor can be achieved.

Recall the theory of Sect. \ref{seccyl}: the word labelling each
cylinder of length $k$ was continued periodically to length $n$ to
single out a cylinder of length $n$ and return time $k$.
Geometrically, for the kind of maps studied in Sect.
\ref{secexpa}, each cylinder $\sigma$ of length $k$ contains a
periodic point of the map, $x_\sigma$, of period $k$. Balls of
radius $\veps$ centered at a point $x$ located in the vicinity of
such fixed point have a non-empty intersection with their $k$-th
iterate if the distance between $x$ and $x_\sigma$ is less than
$\frac{s_k+1}{2(s_k-1)} \veps$, where $s_k$ is the derivative of
$T^k$ at the fixed point $x_\sigma$. Observe that $s_k$ grows
geometrically as $k$ grows, so that when $k$ is large
$\frac{s_k+1}{s_k-1} \simeq 1$. In conclusion, all points $x$ in
the interval $B_{\veps/2}(x_\sigma)$ are such that their
$\veps$-neighborhood returns after time $k$: $T^k(B_\veps(x)) \cap
B_\veps(x) \neq \emptyset$. Of course, not all of these intervals,
labelled by $\sigma$, are disjoint among themselves, both with the
same and different length of $\sigma$. Therefore, two conditions
are to be met to assess a genuine first--return.

We may approximately assume that the first condition (non-overlapping of
the interval $B_{\veps/2}(x_\sigma)$ with other intervals associated with
the same period, $|\sigma|=k$), is met when $\frac{s_k+1}{s_k-1} \veps$ is
less than the geometrical size, $\ell(C_\sigma)$, of the cylinder that
contains the fixed point. More simply, because of our previous
observation, we may just require $\veps < \ell(C_\sigma)$. Let again
$\Sigma_k :=\{0,\ldots,M-1\}^k$ be the set of all words of length $k$.
Within this set we therefore define the subset
\begin{equation}\label{idea11}
  L_{\veps,k} := \{ \sigma \in \Sigma_k \mbox{ s.t. }
  \veps < \ell(C_\sigma) \}.
\end{equation}

The second condition (non-overlapping of $B_{\veps/2}(x_\sigma)$
with intervals of smaller periods $|\sigma|$) is more subtle, and
can be resolved by considering, among all fixed points $x_\sigma$
of period $k$, only the primitively periodic ones. We so define
the set $W_k$:
\begin{equation}\label{idea12}
  W_k := \{ \sigma \in \Sigma_k \mbox{ s.t.  there is no } j < k
 \mbox{ s.t. } \sigma
 \mbox{ is  periodic of period } j \}.
\end{equation}

Summing up, we can write
\begin{equation}\label{idea13c}
 p(\veps,k) =
 \sum_{\sigma \in L_{\veps,k} \cap W_k}
   \mu( B_\veps(x_\sigma) ),
\end{equation}
where $x_\sigma$ is the periodic point in the cylinder $C_\sigma$.
This last expression can be further simplified, using a similar
approximation to that employed in Sect. \ref{seccyl}. In fact, we
can write $\mu( B_\veps(x_\sigma) ) \simeq
\veps^{\alpha_\mu(x_\sigma)}$, where $\alpha_\mu(x_\sigma)$ is the
local dimension of the measure $\mu$ at $x_\sigma$. This last
quantity can then be extrapolated from the measure and the
geometric length of the cylinder $C_\sigma$: $\alpha_\mu(x_\sigma)
\simeq \log(\mu(C_\sigma))/\log(\ell(C_\sigma))$. As a
consequence, eq. (\ref{idea13c}) becomes:
\begin{equation}\label{idea14}
 p(\veps,k) =
 \sum_{\sigma \in L_{\veps,k} \cap W_k}
   \veps^{\log(\mu(C_\sigma))/\log(\ell(C_\sigma))}.
\end{equation}

\begin{figure}[ht]
\includegraphics[width=8cm,height=12cm,angle=270]{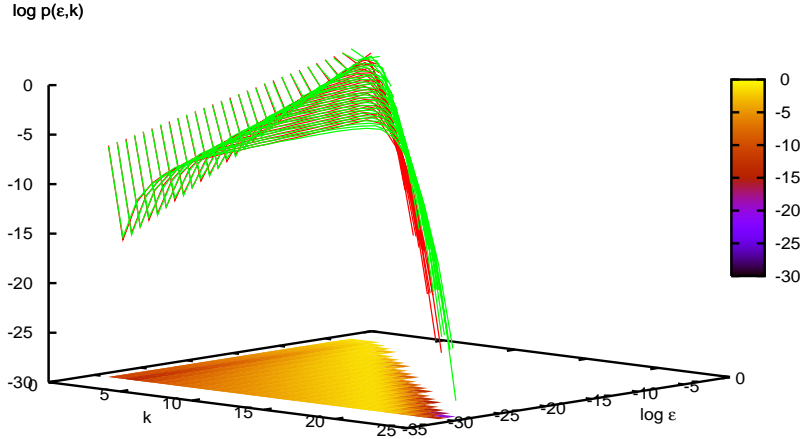}
\caption{Distribution function $p(\veps,k)$ (green lines) from the
original data in Fig. \ref{figr5-2} and approximation from formula
 eq. (\ref{idea14}) (red lines)}
\label{figr1-1}
\end{figure}

We have compared numerically the function $p(\veps,k)$ for the case of
Fig. \ref{figr5-2} of the previous section and its approximation, eq.
(\ref{idea14}), in Fig. \ref{figr1-1}. Agreement is rather satisfactory.

Eq. (\ref{idea14}) can also be written
\begin{equation}\label{idea15}
 p(\veps,k) =
 \sum_{\sigma \in L_{\veps,k} \cap W_k}
   \mu(C_\sigma)^{\log(\veps)/\log(\ell(C_\sigma))}.
\end{equation}
This last equation is particularly meaningful. Observe first that
this generalizes eqs. (\ref{idea1}) and following and can be used
to the same scope. Secondly, put $\lambda_\sigma := -
\log(\ell(C_\sigma))/|\sigma|$. Then, $\mu$ almost surely, when
$|\sigma|$ tends to infinity, $\lambda_\sigma$ converges to the
Lyapunov exponent, $\lambda$, of the measure $\mu$. Therefore,
\begin{equation}\label{idea16}
 p(\veps,k) =
 \sum_{\sigma \in L_{\veps,k} \cap W_k}
   \mu(C_\sigma)^{-\log(\veps)/ \lambda |\sigma|}.
\end{equation}
Suppose now to choose $\veps$ and $|\sigma|$ such that $- \delta \log
\veps = |\sigma| \lambda$, with $0<\delta<1$. This choice has two effects.
First, the exponent in the previous equation becomes $\frac{1}{\delta}$.
Second, the measure of the set $L_{\veps,k}$ tends to one, when $\veps$
tends to zero, because $\mu$ almost surely $-
\log(\ell(C_\sigma))/|\sigma|$ tends to $\lambda$, so that almost surely
$\ell(C_\sigma) > \veps$. Hence,
\begin{equation}\label{idea17}
 p(\veps,-\frac{\delta \log \veps}{\lambda}) \simeq
 \sum_{\sigma \in W_k}
   \mu(C_\sigma)^{\frac{1}{\delta}}.
\end{equation}
If we now let $k$ to be a prime number, the set $W_k$ contains all words
except the ``fixed points'' $\sigma_i=a$, for $i=0,\ldots,k-1$, where $a
\in \{0,\ldots,M-1\}$. In general, one should also subtract all words of
shorter periods that divide $k$, as done above. Call this set of words
$F_k$. Then,
\begin{equation}\label{idea18}
 p(\veps,-\frac{\delta \log \veps}{\lambda}) \simeq
 \sum_{\sigma \in \Sigma_k}
   \mu(C_\sigma)^{\frac{1}{\delta}} - \sum_{\sigma \in F_k}
   \mu(C_\sigma)^{\frac{1}{\delta}}.
\end{equation}
It could be shown that, in systems with sufficiently fast decay of
correlations like Bernoulli or Markov, the first term is dominant in the
limit, so that discarding the second when taking logarithms and dividing
by $\log \veps$, one gets
\begin{equation}\label{idea19}
  \frac{1}{\log \veps}\log p(\veps,-\frac{\delta \log(\veps)}{\lambda})
   = \frac{1}{\log \veps} \log (\sum_{\sigma \in \Sigma_k}
   \mu(C_\sigma)^{\frac{1}{\delta}})=
\frac{\delta}{\lambda} \frac{1}{k} \log (\sum_{\sigma \in \Sigma_k}
   \mu(C_\sigma)^{\frac{1}{\delta}}).
\end{equation}
Taking the limit, we obtain a new verification of eq.
(\ref{idea6c}):
\begin{equation}\label{idea6c2}
  \lim_{\veps \rightarrow 0}
  \frac{1}{\log \veps}\log p(\veps,-\frac{\delta \log(\veps)}{\lambda})
   =
  \frac{(1-\delta)}{\lambda} H (\frac{1}{\delta} - 1).
\end{equation}
In the next section, we shall see a generalization of this equation.

\section{Linear automorphisms of the two-dimensional torus}
\label{seccat}

\begin{figure}[ht]
\includegraphics[width=8cm,height=12cm,angle=270]{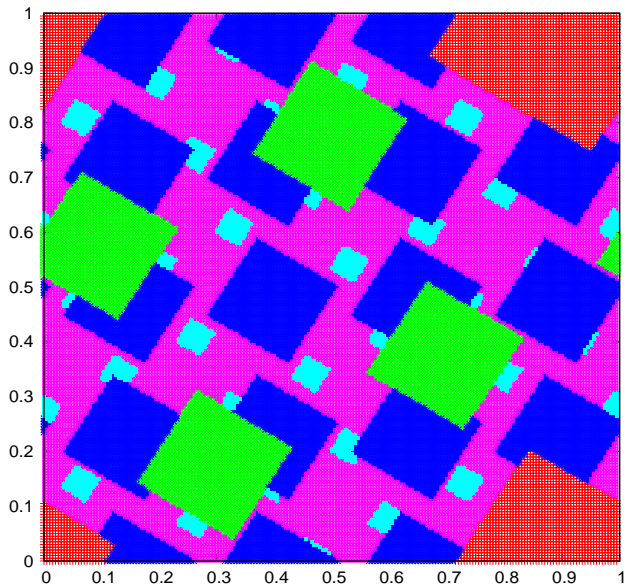}
\caption{Initial conditions $x$ in the two-dimensional torus color
coded according to the return time of the ball $B_\veps(x)$ of
radius $\veps=0.15$ for the Arnol'd cat map. ($\tau=1$ red, 2
green, 3 blue, 4 magenta, 5 light blue)} \label{figmap1-3}
\end{figure}

The general framework presented in the last section can be easily extended
to treat the case of linear automorphisms of the two-dimensional torus, of
which the Arnol'd cat map is the most celebrated example. For convenience,
we choose a metric in the torus such that balls of radius $\veps$ are
euclidean squares of side $2 \veps$ with sides oriented along the stable
and unstable directions and for simplicity we consider the case when these
directions are orthogonal. Then, one can easily show that around any fixed
point of the $k$-th iteration of the map there exists a rectangle, with
sides oriented in the stable and unstable directions, of points $x$ whose
$\veps$-balls intersect their image after $k$ iterations. Letting
$\lambda_-$ and $\lambda_+$ the (increasingly ordered) eigenvalues of $T$,
the sides of this rectangle have length $\veps(1+\frac{2
\lambda_-^k}{\lambda_-^k-1})$ and $\veps(1+\frac{2}{\lambda_+^k-1})$. As
it turns out, for the Arnol'd cat and other area-preserving maps, these
quantities are equal (since $\lambda_+\lambda_-=1$) and the rectangle of
initial conditions just described is a square. Figure \ref{figmap1-3}
draws these squares at a fixed value of $\veps$.

\begin{figure}[ht]
\includegraphics[width=8cm,height=12cm,angle=270]{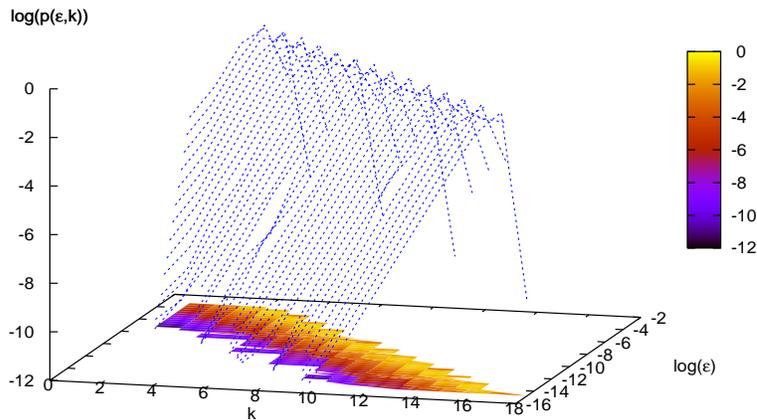}
\caption{Distribution function $p(\veps,k)$ for the toral
automorphism associated with the matrix $(1,2;2,5)$.}
\label{figmap2-4}
\end{figure}

We can then repeat a two-dimensional generalization of the arguments of
the previous section. This we will do elsewhere, but we will provide the
result below. In fact, an even simpler argument can be sketched. We have
seen above that a square of area $\veps^2(1+\frac{2}{\lambda_+^k-1})^2$
exists at each fixed point of $T^k$ and is characterized by return times
$k$, or less. This area quickly becomes $\veps^2$ to a good approximation.
Moreover, the number of periodic points of $T^k$ grows like $\lambda_+^k$.
Then, when $\veps$ is ``small'' with respect to $k$, neglecting all other
considerations, we can write
\begin{equation}\label{cat1}
  p(\veps,k) \simeq \veps^2 \lambda_+^k.
\end{equation}
Of course, we have to make precise what we mean by ``small''. This
is when $\veps^2 \lambda_+^k \leq 1$. Equality holds for $k_\veps
= -\frac{2 \log \veps}{\log(\lambda_+)}$. Indeed, following the
results reported in Sect. \ref{secrev},
$\frac{2}{\log(\lambda_+)}$ is the almost sure limit of
$\frac{\tau({B_\veps(x)})}{-\log \veps}$, since it coincides with
$ \frac{1}{\log(\lambda_+)}-\frac{1}{\log(\lambda_-)}$, see eq.
(\ref{duedim}). Moreover, in this case $D_1(\mu)=2$, the invariant
measure being the Lebesgue measure and the entropy $h_\mu$ is
equal to the Lyapunov exponent $\lambda=\log(\lambda_+)$. Figure
\ref{figmap2-4} draws the distribution function $p(\veps,k)$ for a
different toral automorphism, just chosen to increase variety:
that associated with the matrix $(1,2;2,5)$. The logarithmically
flat approximation in eq. (\ref{cat1}) fits the data almost
perfectly in the region $\veps^2 \lambda_+^k \leq 1$.

If we now turn our consideration to the line $k= \delta k_\veps$ in the
$(k,\log \veps)$ plane, with $0<\delta \leq 1$, we can prove that the
quantity $p(\veps,\delta k_\veps)$ verifies in this two-dimensional case
the analogue of eq. (\ref{idea6b}):
\begin{equation}\label{cat4}
  \lim_{\veps \rightarrow 0}
  \frac{1}{\log \veps}\log p(\veps,-\frac{\delta \log(\veps)}{\lambda/2})
   =
  \frac{(1-\delta)}{\lambda/2} H (\frac{1}{\delta} - 1).
\end{equation}
The detailed proof, obtained along the lines of Sect.
\ref{secexpa2} will be reported elsewhere. We simply compute here
the two sides of the equality (\ref{cat4}), showing that they are
equal. From eq. (\ref{cat1}) we can compute the l.h.s., obtaining
\begin{equation}\label{cat5}
  \lim_{\veps \rightarrow 0}
  \frac{1}{\log \veps}\log p(\veps,-\frac{\delta \log(\veps)}{\lambda/2})
   =
  2 (1-{\delta}).
\end{equation}
On the other hand, the R\'enyi entropies for the Lebesgue measure
and the Arnol'd cat dynamics are all equal to $\lambda
=\log(\lambda_+)$, so that also the l.h.s. of eq. (\ref{cat4}) is
equal to $2(1-\delta)$.

We conclude this section by taking inspiration from eq.
(\ref{cat4}) to put forward a conjecture. We believe that, letting
$\eta$ be the almost sure limit of
$\frac{\tau({B_\veps(x)})}{-\log \veps}$, see eq. (\ref{lmtb}),
and letting $k_\veps = - \eta \log \veps$ as before, under
sufficient hypotheses of mixing, one has the following asymptotic
behavior:
\begin{equation}\label{catconj}
  \lim_{\veps \rightarrow 0}
  \frac{1}{\log \veps}\log p(\veps,\delta k_\veps)
   =
  \frac{(1-\delta)}{\eta} H (\frac{1}{\delta} - 1).
\end{equation}
Quite evidently, further investigation is required to confirm this
conjecture. We can now turn to conclusions.

\section{Conclusions}
We have studied in this paper various aspects of the statistics of
return times of sets in dynamical systems.

We have first reviewed known results for symbolic $\psi$-mixing
systems, that link return times and R\'enyi entropies. We have
established new ``counting rules'', embodied in the sets
${\mathcal W}_{n,k}$ and in the lemma of eq. (\ref{idea10}) that
have permitted us to explain the slow convergence of the
quantities studied in previous works. At the same time, these
results have lead to the definition of a new ``partition
function'' $Z_p(\delta,n)$, eq. (\ref{idea15a}), that best
achieves the goal of extracting R\'enyi entropies from return
times statistics.

When considering return times for balls, we have established a
general relation holding for one-dimensional expanding maps, eq.
(\ref{idea6c}), that links the asymptotic of return times with
R\'enyi entropies {\em and} the Lyapunov exponent. This relation
has been obtained developing two different approaches. The former
is a quantitative comparison between balls and dynamical cylinders
especially developed for this case. The second is a more general
argument that well describes the full behavior of the statistics
$p(\veps,k)$, comprised in eq. (\ref{idea14}).

We have finally considered linear automorphisms of the two
dimensional torus, like the Arnol'd cat map, for which a ``quick
and dirty'' analysis is capable of describing the correct behavior
of the distribution function $p(\veps,k)$. This has permitted us
to write the formula in eq. (\ref{cat4}) that links return times
statistics and R\'enyi entropies with $\eta$, the almost-sure
value of the limit in eq. (\ref{lmtb}). This formula is an
extension of that obtained for one-dimensional systems and we have
conjectured that it should hold in much more general situations
than the one presented in this paper.

As stated in the Introduction, the character of this paper is
tailored to the audience expected for this volume, that comprises
both specialists in dynamical systems and in other disciplines. We
have therefore tried to present our results in the most
transparent form, while renouncing at times to full rigor in favor
of clarity. We are nonetheless convinced that most of the theory
developed here touches upon new ideas and approaches, and presents
more than valuable hints to where a rigorous treatment will be
developed, as we plan to do in forecoming publications.

\end{document}